\documentclass{article}
\usepackage{amsfonts}

\newcommand{\bls}[1]{\renewcommand{\baselinestretch}{#1}}
\def\noi{\noindent}

\makeatletter
\renewcommand{\section}{\@startsection{section}{1}{0pt}%
        {-3.5ex plus -1ex minus -.2ex}{2.3ex plus .2ex}%
        {\large\bf\protect\raggedright}}

\renewcommand{\subsection}{\@startsection{subsection}{2}{0pt}%
        {-3ex plus -1ex minus -.2ex}{1.4ex plus .2ex}%
        {\normalsize\bf\protect\raggedright}}
\renewcommand{\paragraph}{\@startsection{paragraph}{4}{0pt}%
        {2ex plus -.5ex minus -.2ex}{-1em}{\normalsize\bf}}

\renewcommand{\thesubsubsection}%
        {\arabic{section}.\arabic{subsection}.\arabic{subsubsection}.}

\renewcommand{\@oddhead}{\raisebox{0pt}[\headheight][0pt]{%
   \vbox{\hbox to\textwidth{\rightmark \hfil \rm \thepage \strut}\hrule}}}

\makeatother

\newcommand{\Title}[1]{\noi {\Large #1} \\}

\newcommand{\Authors}[4]{\noi
        {\large\bf #1\dag\ #2\ddag}\medskip\begin{description}
        \item[\dag]{\it #3} \item[\ddag]{\it #4}\end{description}}

\newcommand{\Abstract}[1]{\vskip 2mm \begin{center}
        \parbox{16.4cm}{\small\noi #1} \end{center}\medskip}

\newcommand{\sect}[1]{Sec.\,#1}

\def\nqq{\hspace*{-2em}}
\def\nhq{\hspace*{-0.5em}}

\def\cm{\hspace*{1cm}}

\def\para{\paragraph}
\newcommand{\Theorem}[2]{\medskip\noi {\bf #1. \ }{\it #2}\medskip}

\def\eq{Eq.\,}
\def\eqs{Eqs.\,}
\def\beq{\begin{equation}}
\def\eeq{\end{equation}}
\def\bear{\begin{eqnarray}}
\def\al{&\nhq}
\def\lal{&&\nqq {}}               
\def\bearr{\begin{eqnarray} \lal}
\def\ear{\end{eqnarray}}
\def\earn{\nonumber \end{eqnarray}}
\def\dst{\displaystyle}
\def\tst{\textstyle}

\newcommand{\fract}[2]{{\tst\frac{#1}{#2}}}
\def\nn{\nonumber\\ {}}
\def\nnv{\nonumber\\[5pt] }
\def\nnn{\nonumber\\ \lal }

\def\yy{\\[5pt] {}}

\def\eql{\al =\al}
\def\eps{\varepsilon}

\def\eqdef{\stackrel{\rm def}=}
\def\e{{\,\rm e}}
\def\d{\partial}

\def\sign{\mathop{\rm sign}\nolimits}
\def\diag{\mathop{\rm diag}\nolimits}

\def\const{{\rm const}}
\def\Half{{\dst\frac{1}{2}}}
\def\half{{\tst\frac{1}{2}}}
\def\then{\ \Rightarrow\ }

\def\Jl#1#2{{\it #1\/} {\bf #2},\ }

\def\NP#1 {\Jl{Nucl. Phys.}{#1}}
\def\PRD#1 {\Jl{Phys. Rev.}{D#1}}
\def\DAN#1 {\Jl{Dokl. AN SSSR}{#1}}
\def\JETF#1 {\Jl{Zh. Eksp. Teor. Fiz.}{#1}}
\def\JMP#1 {\Jl{J. Math. Phys.}{#1}}
\def\PLA#1 {\Jl{Phys. Lett.}{A#1}}
\def\PLB#1 {\Jl{Phys. Lett.}{B#1}}
\def\PRL#1 {\Jl{Phys. Rev. Lett.}{#1}}

\def\GR{general relativity}
\def\cyl{cylindrically symmetric}
\def\sol{soliton-like}

\def\R{{\mathbb R}}
\def\cU{{\cal U}}
\def\thd{\fract 13}
\def\mn{_{\mu\nu}}
\def\MN{^{\mu\nu}}
\def\mN{_\mu^\nu}
\def\nM{_\nu^\mu}
\def\kappa{\mbox{\sl \ae}}

\def\xa{x_{\rm ax}}
\def\xh{x_{\rm hor}}
\def\xinf{x_{\infty}}
\def\af{{\alpha_\varphi}}
\def\aff{{\alpha_{\varphi\varphi}}}

\bls{1.02}
\begin{document}
\thispagestyle{empty}

\rightline{\large\bf gr-qc/0101086}
\medskip

\Title
{CYLINDRICALLY SYMMETRIC SOLITONS \yy
 WITH NONLINEAR SELF-GRAVITATING SCALAR FIELDS}

\Authors{K.A. Bronnikov} {and G.N. Shikin}
{Centre for Gravitation and Fundam. Metrology, VNIIMS,
        3-1 M. Ulyanovoy St., Moscow 117313, Russia;\\
Institute of Gravitation and Cosmology, PFUR,
        6 Miklukho-Maklaya St., Moscow 117198, Russia}
{Dept. of Theoretical Physics, PFUR,
        6 Miklukho-Maklaya St., Moscow 117198, Russia}

\Abstract
{Static, \cyl\ solutions to nonlinear scalar-Einstein equations are
considered. Regularity conditions on the symmetry axis and flat or string
asymptotic conditions are formulated in order to select \sol\ solutions.
Some non-existence theorems are proved, in particular, theorems
asserting (i) the absence of black-hole and wormhole-like \cyl\ solutions
for any static scalar fields minimally coupled to gravity and (ii) the
absence of solutions with a regular axis for scalar fields with the
Lagrangian $L=F(I)$, $I=\varphi^\alpha \varphi_\alpha$, for any function
$F(I)$ possessing a correct weak field limit. Exact solutions for scalar
fields with an arbitrary potential function $V(\varphi)$ are obtained by
quadratures and are expressed in a parametric form in a few ways, where the
parameter may be either the coordinate $x$, or the $\varphi$ field itself,
or one of the metric coefficients. It is shown that \sol\ solutions exist
only with $V(\varphi)$ having a variable sign.  Some explicit examples of
the solutions (including a \sol\ one) and their flat-space limit are
discussed.  }

\section{Introduction}

    The concept of solitons and particle-like configurations in nonlinear
    field theory has appeared as one of the approaches aimed at avoiding the
    well-known difficulties of the theories describing particles as
    mathematical points.  In this approach, a hope to create a
    divergence-free particle theory was connected with a search for and
    studies of exact, regular, localized solutions to classical nonlinear
    field equations, able to describe the complicated spatial structure of
    particles observed in the experiment \cite{c1}.  It was evident that
    nonlinearity should be necessarily included in the field equations in
    order to describe field interaction, irrespective of the divergence
    problem. In other words, nonlinearity is not only one of possible ways
    of removing the difficulties of the theory, but a reflection of real
    field properties.

    Nowadays the problem of infinities is mostly discussed and solved in
    the context of numerous versions of string theory, which is known to be
    most promising on the main trend of modern theoretical physics,
    unification of the four interactions. Meanwhile, it is string theory
    (along with gauge field theory) that has created a new burst of
    interest in classical nonlinear field theories.  On the one hand, there
    naturally appear various scalar fields with nonlinear potentials, on
    the other, some models of string theory create in their low-energy
    limits such theories as the Born-Infeld nonlinear electrodynamics or
    its non-Abelian modifications \cite{BI}. And, as previously, solitonic
    solutions are of utmost importance in any such theory.

    Many papers, devoted to soliton-like solutions to nonlinear field
    equations, disregard the self-gravity of the field system under study,
    although its inclusion is of great interest since the gravitational
    field is intrinsically nonlinear, universal and cannot be shielded;
    moreover, the inclusion of self-gravity can drastically change the
    properties of solutions to nonlinear field equations and even their
    existence conditions \cite{c2}.

    Of greatest physical interest are evidently spherically symmetric (or
    more general axially symmetric) solutions able to describe localized
    objects in real three-dimensional space. However, some problems
    necessitate studies of two-dimensional, or \cyl\ solutions, localized
    in a neighbourhood of the symmetry axis, the so-called vortices or
    string-like solutions \cite{c3, c4}. Such solutions can both describe
    certain realistic objects like superconducting fibres (fluxons)
    \cite{c5} or light beams \cite{c6} and serve as reasonable
    approximations for toroidal structures when a torus of large radius is
    replaced by a closed string \cite{c7}. In the case of self-gravitating
    configurations, a natural application of \sol\ structures is the
    description of cosmic strings beyond the approximation treating them as
    simple conical singularities \cite{s1}--\cite{s6}.

    In this paper we discuss static, \cyl, \sol\ configurations of
    nonlinear scalar fields with various Lagrangians in \GR. Such scalar
    fields can be of any origin: Higgs fields, dilatons, inflatons, etc.
    The term ``\sol'' will here mean a globally isolated regular field
    configuration seen by a distant observer as a gravitating cylinder or a
    cosmic string.

    In \sect 2 we formulate the regularity and asymptotic conditions
    to be satisfied by the sought solutions.  In \sect 3 we prove some
    statements showing which kinds of \sol\ solutions cannot exist.
    It turns out, in particular, that configurations of scalar fields with
    Lagrangians of the form $L=F(I)$, $I= \varphi^\alpha \varphi_\alpha$
    (for which the field equations are solved by quadratures)
    cannot have a regular axis, whatever is the function $F(I)$.
    It is also shown that nonlinear scalar fields cannot lead to \cyl\
    analogues of black-hole and wormhole solutions.
    \sect 4 is devoted to scalar fields with nonlinearities in the form of
    an arbitrary potential function $V(\varphi)$. It is shown, in
    particular, that \sol\ solutions with this kind of nonlinearity can
    exist only if $V(\varphi)$ has a variable sign. Treating the potential
    $V(\varphi)$ as one of unknown functions, so that the set of field
    equations is underdetermined, we describe four ways of obtaining exact
    solutions by quadratures. The first way
    requires specifying the function $\alpha(x)$, the second ---
    $V(\alpha)$ (where $\e^{2\alpha}$ is one of the metric coefficients and
    $x$ is the radial coordinate). In the third approach one should specify
    the function $\varphi(\alpha)$ while the fourth one starts with a given
    $\alpha(\varphi)$.  The nontrivial nature of the flat-space limit of
    these solutions is discussed in \sect 5: when the gravitational
    constant $\kappa\to 0$, this is only a necessary rather than
    sufficient condition for passing to a flat-space solution. In \sect 6
    we consider three examples of solutions with different $V(\varphi)$.
    The first one is the Liouville nonlinearity for which exact solutions
    are obtained directly, but among them there is no \sol\ one. The second
    one illustrates a connection between self-gravitating and flat-space
    solutions. The third one is an example of a \sol\ solution, which
    is given in a parametric form, in terms of elliptic functions.

\section 
    {Regularity and asymptotic conditions}

    Let us write down the metric without fixing the radial coordinate $x$:
\beq                                                         \label{e2}
    ds^2 = \e^{2\gamma} dt^2 - \e^{2\alpha} dx^2
    				 -\e^{2\xi} dz^2 - \e^{2\beta} d\phi^2
\eeq
    where $\alpha,\beta,\gamma,\xi$ are functions of $x$; $z\in\R$ and
    $\phi \in [0,2\pi)$ are the longitudinal and azimuthal coordinates,
    respectively.

    We will try to select \sol\ configurations from the whole set of
    solutions. As mentioned above, this will imply two requirements:
    the existence of a spatial asymptotic from which our system is seen as
    an isolated \cyl\ source of gravity or a cosmic string, and global
    regularity of the space-time and the fields.
    If the $(x,\,\varphi)$ surfaces are simply connected, the global
    regularity condition actually reduces to that of regularity on the
    symmetry axis. Another opportunity, a wormhole-like topology of the
    $(x,\,\phi)$ surfaces, will be discussed in \sect 3.

\subsection 
    {Regularity on the axis}

    The regularity conditions on an axis, i.e. at a value $\xa$ of $x$ such
    that $\e^{\beta}\to 0$, include the finiteness requirement for the
    algebraic curvature invariants and the condition
\beq
    |\beta'| \e^{\beta-\alpha}\to 1 			    \label{con}
\eeq
    (where the prime denotes $d/dx$), expressing a correct relation between
    infinitesimal circumferences and radii, in other words, the absence of a
    conical singularity.  Among the curvature invariants it is sufficient to
    deal with the Kretschmann scalar
    $K = R^{\mu\nu\rho\sigma}R_{\mu\nu\rho\sigma}$ which, for the metric
    (\ref{e2}), is a sum of squared components of all nonzero Riemann tensor
    components $R\MN{}_{\rho\sigma}$:
\bear
    K \eql 4 \sum_{i=1}^{6} K_i^2; \nnv
    K_1 \eql R^{01}{}_{01}
    		=-\e^{-\alpha-\gamma}(\gamma'\e^{\gamma-\alpha})',
  \cm
    K_2 = R^{02}{}_{02} = -\e^{-2\alpha}\gamma'\xi',
  \nn
    K_3 \eql R^{03}{}_{03} = -\e^{-2\alpha}\beta'\gamma',
  \cm\cm\
    K_4 = R^{12}{}_{12} = -\e^{-\alpha-\xi}(\xi'\e^{\xi-\alpha})',
  \nn
    K_5 \eql R^{13}{}_{13} = -\e^{-\alpha-\beta}(\beta'\e^{\beta-\alpha})',
  \cm
    K_6 = R^{23}{}_{23} = -\e^{-2\alpha}\beta'\xi'              \label{Kr}
\ear
    For $K < \infty$ it is thus necessary and sufficient that all $|K_i| <
    \infty$, and this in turn guarantees that all algebraic invariants of
    the Riemann tensor will be finite.  Note that all $K_i$, as well as the
    condition (\ref{con}), are invariant under reparametrization of $x$.

    From (\ref{con}) and $\beta\to -\infty$ it follows that $K_3$ and $K_6$
    are finite if and only if
\beq
     \gamma'\e^{-\alpha} = O(\e^{\beta}),\cm                    \label{reg1}
     \xi'\e^{-\alpha} = O(\e^{\beta}),
\eeq
    and consequently $K_2 = O(\e^{2\beta})\to 0$. Here and henceforth
    the symbol $O(f)$ denotes a quantity either of the same
    order of magnitude as $f$ in a certain limit, or smaller, while
    the symbol $\sim$ connects quantities of the same order of magnitude.

    It is easily shown that the conditions (\ref{reg1}) can
    only hold if $\gamma$ and $\xi$ take finite values on the axis.

    The remaining quantities $K_1$, $K_4$, $K_5$ are better dealt with using
    specific coordinates. Let us choose the harmonic $x$ coordinate, such
    that
\beq
    \alpha = \beta + \gamma+ \xi.                              \label{harm}
\eeq
    Then, as $\beta\to -\infty$, $\alpha=\beta + O(1)$, and, since by
    (\ref{con}) $\beta'\sim 1$, it is evident that a regular axis can only
    occur at $x= \xa = \pm \infty$. Choosing $\xa = -\infty$, one can write:
\beq
	\beta = cx (1 + o(1)), \cm
	c = \e^{\gamma+\xi}\Big|_{x\to -\infty}  = \const >0,  \label{reg2}
\eeq
    and, as follows from (\ref{reg1}),
\beq
	\gamma' = O(\e^{2cx}), \cm \xi' = O(\e^{2cx}).         \label{reg3}
\eeq
    One can now check that under these conditions $K_1$ and $K_4$ are finite
    on the axis, while the finiteness of $K_5$ requires
    $\beta'' = O(\e^{2cx})$. This means that the condition (\ref{con})
    should be strengthened, and in a reparametrization-invariant form we
    have
\beq
    |\beta'| \e^{\beta-\alpha}= 1 + O(\e^{2\beta}).   	       \label{con1}
\eeq

    Thus to provide a regular axis it is necessary and sufficient to
    require the validity of (\ref{reg1}) and (\ref{con1}) as $x\to \xa$.
    The same conditions in terms of the conventional radial coordinate
    $r = \e^\beta$ read
\beq
     \e^\alpha = 1 + O(r^2); \cm
     \gamma = \gamma_{\rm ax} + O(r^2); \cm
     \xi = \xi_{\rm ax}+O(r^2) \cm {\rm as} \quad r\to 0.      \label{ax-r}
\eeq

    Another useful necessary condition for regularity follows
    from the Einstein equations. At points of a regular axis, as at any
    regular space-time point, the curvature invariants $R$ and
    $R\mn R\MN$ should be finite. Since the Ricci tensor for the metric
    (\ref{e2}) is diagonal, the invariant $R\mn R\MN \equiv R\mN R\nM$ is a
    sum of squares, hence each component $R_\mu^\mu$ (no summing) is finite
    at a regular space-time point. Then, by virtue of the Einstein
    equations, each component of the EMT $T\mN$ is finite as well:
\beq
	|T\mN| < \infty.                                     \label{Tfin}
\eeq
    Thus, requiring only the regularity of the geometry, we obtain,
    as its necessary condition, the finiteness of all EMT components.
    This is true not only for the present case, but always when $R\mN$
    is diagonal.

\subsection  
    {Regular (flat and string) asymptotics}

    We will be only concerned with isolated \cyl\ configurations and
    therefore do not consider solutions having asymptotics of cosmological
    nature, such as closed models like the Melvin magnetic universe or
    those with (anti-)de Sitter asymptotics which should appear where the
    EMT behaves like a cosmological constant. We shall instead require the
    existence of a spatial infinity, i.e., such $x=\xinf$ that $\beta \to
    \infty$, where the metric is either flat, or corresponds to the
    gravitational field of a cosmic string.

    This means that, first, as $x\to\xinf$,
    a correct behavior of clocks and rulers requires
    $|\gamma| < \infty$ and $|\xi| < \infty$ as $x\to \xinf$, or, choosing
    proper scales along the $t$ and $z$ axes, one can write
\beq
     \gamma \to 0, \cm \xi \to 0 \cm {\rm as} \quad x\to\xinf. \label{as1}
\eeq

    Second, at the asymptotic the condition (\ref{con}) should be replaced by
    a more general one,
\beq
    |\beta'| \e^{\beta-\alpha}\to 1 - \mu, \cm \mu = \const < 1 \cm
    		{\rm as} \quad x\to\xinf,                 \label{defect}
\eeq
    so that the circumference to radius ratio for the circles $x=\const$,
    $z=\const$ tends to $2\pi(1-\mu)$ instead of $2\pi$. In this case the
    space-time is locally flat but behaves asymptotically as if it were flat
    everywhere but on the axis, where a conical singularity is located,
    creating the angular defect $\mu$. In other words, under the asymptotic
    conditions (\ref{as1}), (\ref{defect}), $\mu>0$, a \sol\ solution can
    simulate a cosmic string. A flat asymptotic takes place if $\mu=0$.

    In what follows we will use the words ``{\it regular asymptotic\/}'' in
    the sense ``{\it flat or string asymptotic\/}''.

    Third, the curvature tensor should vanish at the asymptotic, and, by
    virtue of the Einstein equations, all the EMT components must decay
    quickly enough.  It can be easily checked that the conditions
    (\ref{as1}) and (\ref{defect}) automatically imply that all $K_i =
    o(\e^{-2\beta})$ where $K_i$ are defined in (\ref{Kr}). Consequently the
    same decay rate at a regular asymptotic takes place in all components of
    $T\mN$, and one can verify, in particular, that the total material field
    energy per unit length along the $z$ axis is finite:
\beq
    \int T^0_0 \sqrt{-{}^3g}\, d^3x =
    \int T^0_0 \e^{\alpha+\beta+\xi}\,dx\,dz\,d\phi < \infty \label{E_field}
\eeq
    where integration in $z$ covers a unit interval. A similar condition in
    flat-space field theory is used as a criterion of field energy being
    localized around the symmetry axis, which is one of the requirements to
    solitonic solutions. The set of asymptotic regularity requirements
    (\ref{as1}), (\ref{defect}) for self-gravitating solutions is thus much
    stronger than (\ref{E_field}) and contains the latter as a corollary.

    It should also be noted that even the vacuum \cyl\
    solution has in general no regular asymptotic; this is,
    physically, due to an infinite total mass of an infinitely long source
    cylinder. The only vacuum solution with a regular asymptotic is
    described by flat space-time metric, maybe with a conical singularity
    on the axis. So our requirement means that the \sol\ solutions sought
    for should behave asymptotically just as this particular vacuum
    solution.

    A static, linear, massless, minimally coupled scalar field also cannot
    provide a regular asymptotic.  Indeed, e.g., in the coordinates
    (\ref{harm}) the field equation reads $\varphi''=0$, and its nontrivial
    solution, $\varphi'=\const\ne 0$, leads to
    $T^0_0 \sim \e^{-2\alpha} \sim \e^{-2\beta}$ as $x\to\infty$ if one
    requires (\ref{as1}) and (\ref{defect}). Then the integral
    (\ref{E_field}) diverges at $x\to \infty$. Thus, unlike spherical
    symmetry (where even linear fields vanish quickly enough), a regular
    \cyl\ asymptotic is only possible due to an essentially nonlinear
    behavior of material fields.

\section{Field equations and non-existence theorems}    

\subsection{Einstein equations and regularity conditions}      

    For the metric (\ref{e2}),
    under the coordinate condition (\ref{harm}),
    we can write down the Einstein equations in the form
\bear                                                        \label{e4}
     \beta'' + \xi'' -\cU \eql -\kappa T_0^0 \e^{2\alpha}, \nn
                      \cU \eql -\kappa T_1^1 \e^{2\alpha}, \nn
    \gamma'' + \xi'' -\cU \eql -\kappa T_2^2 \e^{2\alpha}, \nn
   \beta'' +\gamma'' -\cU \eql -\kappa T_3^3 \e^{2\alpha}
\ear
    where $\cU \eqdef \beta'\gamma' + \beta'\xi' + \gamma'\xi'$,
    and $T\mN$ is the energy-momentum tensor (EMT). For any static scalar
    fields minimally coupled to gravity it has the property of importance
\beq
    T_0^0 = T_2^2 = T_3^3.                                     \label{TTT}
\eeq
    Therefore \eqs (\ref{e4}) combine to give
\beq
    \beta''=\gamma''= \xi'' = \thd\alpha''                     \label{e6a}
\eeq
    where the last equality is due to (\ref{harm}), whence
\bear
		\xi \eql \thd(\alpha-Ax),\nn
	     \gamma \eql \thd(\alpha-Bx),\nn                   \label{e7a}
	      \beta \eql \thd(\alpha+Ax+Bx)
\ear
    where $A$ and $B$ are integration constants and other two constants
    are ruled out by a proper choice of the origin of $x$ and the scale
    along the $z$ axis.

    The function $\beta(x)$ determines the nature of the static space.
    In particular, as discussed above, $\beta\to -\infty$ corresponds to an
    axis, if any; at an asymptotic $\beta\to\infty$.

    In the coordinates (\ref{harm}) the conditions (\ref{con}) or
    (\ref{defect}) can only hold at $x\to\pm\infty$. It is evident from
    (\ref{e7a}) that if one requires either a regular axis (say, at
    $x\to-\infty$) or a regular asymptotic (at $x=+\infty$), the constants
    should satisfy the requirement
\beq
	A = B = N >0.                                        \label{ABN}
\eeq
    So the regular axis and regular asymptotic requirements lead to the same
    relation (\ref{ABN}) for the integration constants, i.e., are
    compatible; this is favourable for the existence of \sol\ solutions.

    Suppose there is a \sol\ solution with a regular axis and a regular
    asymptotic. Then at both ends one has
\beq
       \alpha \approx \beta \approx  Nx,                     \label{abNx}
\eeq
    with the same constant $N$.

    Under the conditions (\ref{as1}) at a regular asymptotic, the constant
    $N$ has a clear geometric meaning. Indeed, according to (\ref{defect})
\beq
	N = 1-\mu                                            \label{Nmu}
\eeq
    where $\mu$ is the angular defect at a string asymptotic.

    On the other hand, comparing (\ref{abNx}) and (\ref{reg2}), one sees
    that $c=N$, so that
\beq
	N = \e^{2\gamma_{\rm ax}} = \e^{2\xi_{\rm ax}}.      \label{Nax}
\eeq
    Thus the angular defect is directly related to the values of $g_{tt}=
    \e^{2\gamma}$ and $g_{zz}=\e^{2\xi}$ on the axis. In particular,
    $g_{00}$ and its gradient determine the course of clocks and the
    gravitational forces applied to test particles at rest, respectively.
    So one can conclude that solitons with a string asymptotic ($\mu >
    0$) have (at least on the average) an attracting gravitational field,
    and photons coming from the axis are redshifted, whereas for solitons
    with a flat asymptotic ($\mu = 0$) both redshifts and forces are
    averaged to zero on the way from the axis to spatial infinity.

    \eq (\ref{abNx}) can be further refined in a way similar to (\ref{ax-r})
    using (\ref{reg3}) and (\ref{con1}), namely, near the axis
    ($x\to -\infty$)
\beq
	\alpha= Nx + O(\e^{2Nx}), \cm
	\beta = Nx + O(\e^{2Nx}).                             \label{ax-x}
\eeq

    At a regular asymptotic ($x\to\infty$) one has, according to
    (\ref{e7a}) and (\ref{abNx}),
\beq
	\alpha= Nx + o(1), \cm
	\beta = Nx + o(1).                                    \label{as-x}
\eeq

\subsection               
	{Non-existence of black-hole, wormhole and hornlike solutions}

    It can be shown that certain types of behavior of the solutions are
    incompatible with scalar fields as sources of geometry.

    An opportunity of interest is the existence of \cyl\ configurations
    similar to black holes (``black strings''), i.e., those with a cylinder
    $x=\xh$ having the properties of a horizon. Some necessary conditions
    for that are: (i) this surface is regular, so that all $K_i$ defined in
    (\ref{Kr}) are finite, (ii) $\e^{\gamma(\xh)}=0$ (a Killing horizon
    for the timelike Killing vector) and (iii) $\xi(\xh)$ and $\beta(\xh)$
    are finite.

    One can easily show that in the coordinates (\ref{harm}) these
    conditions are feasible only as $x\to \pm\infty$; assuming that
    $x=+\infty$ is the asymptotic, we are left with $\xh = -\infty$.
    Then all $K_i$ are finite only if in (\ref{e7a}) $A = -B/2 \ne 0$.
    This is clearly in contrast to (\ref{ABN}) if we require that the same
    solution has a regular asymptotic. We have to conclude:

\Theorem{Proposition 1}
    {Static, \cyl\ black holes with a regular asymptotic cannot exist in
    \GR\ with matter whose EMT satisfies \eq (\ref{TTT}).}

    A nonsingular \cyl\ solution does not necessarily have a regular axis:
    it may contain no axis at all, so that the circularly symmetric
    ($x,\phi$) surfaces have the topology of a cylinder. Such possible
    cases are
\begin{description}
\item [(i)]
    a wormhole-like configuration, which, by definition, possesses
    two spatial infinities connected by a neck, i.e., a regular minimum of
    the function $\beta(x)$;
\item [(ii)]
    a hornlike configuration, where $\beta(x)$ monotonically approaches
    $\beta_{\min}$ as $x$ tends to a certain limiting value $x^*$; the
    space-time is nonsingular if, as $x\to x^*$, the metric coefficients
    $\e^{2\beta(x)}$, $\e^{2\gamma(x)}$ and $\e^{2\xi(x)}$ have finite
    limits while the integral $l = \int \e^\alpha du$ diverges. In
    other words, each ($x,\phi$) surface ends with a regular infinitely long
    tube of finite radius.
\end{description}

    Let us discuss these opportunities for nonlinear scalar fields in \GR.

    Suppose first that there are {\it two regular spatial asymptotics}.
    As before, one of them is at $x\to +\infty$.
    At this asymptotic $\alpha \approx \beta$ and \eq (\ref{ABN}) holds;
    from (\ref{e7a}) one easily finds that $\alpha \approx\beta\approx Nx$.
    Another regular asymptotic might occur at $x \to -\infty$; however,
    since the relation for the integration constants $A=B=N$ still holds, if
    we assume that $\gamma$ and $\xi$ are finite there, we arrive again at
    $\alpha \sim \beta \sim Nx$, but now it means that $\beta\to
    -\infty$, that is, an axis (which can in principle be regular); another
    spatial infinity cannot exist. We arrive at the following result:

\Theorem{Proposition 2}
    {Static, \cyl\ wormholes with two regular asymptotics do not exist in
    \GR\ with matter whose EMT satisfies \eq (\ref{TTT}).}

    If we deny the asymptotic regularity condition but require symmetry of a
    wormhole-like configuration with respect to its neck, then at such a
    neck $\beta' = \gamma' = \xi' =0$, hence $\cU=0$.
    With $\cU=0$, \eqs (\ref{e4}) can be combined to give
\beq
	\beta'' = -\e^{2\alpha}(T^0_0 + T^2_2 - T^3_3)        \label{minB}
\eeq
    On the other hand, a minimum of $\beta$ implies $\beta''>0$ on the
    neck and in its certain neighbourhood where the first-order derivatives
    are small compared with the second-order ones and $\cU=0$ remains a
    valid approximation.  Assuming $T^2_2=T^3_3$ (which is true for scalar
    fields), \eq (\ref{minB}) then means that in the same neighbourhood
    $T_0^0 <0$ (negative energy density). The result is:

\Theorem{Proposition 3}
    {A static, \cyl\ wormhole, symmetric with respect to its neck, cannot
    exist in \GR\ with matter whose EMT satisfies the conditions
    $T^2_2 = T^3_3$ and $T^0_0 \geq 0$.}

\noi{\it Remark.\/} The proof of Prop.\,3 may be readily refined to
include the case that, at the minimum of $\beta$, $\beta'' = 0$ but the
lowest nonzero derivative is even-order and positive. The result will be the
same. One can also observe that, to have $\cU=0$ on the neck, it is
sufficient to require only $\gamma'=0$ {\it or\/} $\xi'= 0$ rather than
both.  The conditions of Prop.\,3 may be accordingly weakened.

    Suppose now that there is a hornlike solution with a regular
    asymptotic. If the latter occurs at $x \to \infty$, then the ``horn''
    $x^* = -\infty$, since, with $\alpha = \beta+\gamma+\xi$ finite,
    the integral $l$ can diverge only at infinite $x$. One then has to
    require in \eq (\ref{e7a}) that $A=B=0$, whereas a regular asymptotic
    requires the validity of (\ref{ABN}). This proves the following:

\Theorem{Proposition 4}
    {Static, \cyl\ hornlike solutions with a regular asymptotic do not
    exist in \GR\ with matter whose EMT satisfies \eq (\ref{TTT}).}

    These restrictions should be taken into account in the further analysis.
    In particular, we shall not seek black-hole, wormhole or
    hornlike solutions.  Consequently, in what follows a {\it \sol\
    configuration\/} will mean a configuration with a regular axis and a
    regular asymptotic.

\subsection  
	{Self-gravitating scalar field with the nonlinearity $L = F(I)$,
	 $I= \varphi^\alpha \varphi_\alpha$}

    Consider a nonlinear scalar field in \GR, described by the
    total Lagrangian
\beq
	L = \frac{R}{2\kappa} + F(I),\cm                     \label{e1}
	 			I= \varphi^\alpha \varphi_\alpha,
\eeq
    where $R$ is the scalar curvature. It is assumed that for weak
    fields ($I\to 0$) the scalar field Lagrangian $F(I)$ behaves like that
    of a linear field: $F =\half I + o(I)$ ({\it a linear weak field
    limit}). The corresponding EMT is
\bear
     T\mN \eql 2 \frac{dF}{dI}\varphi_{,\mu}\varphi^{,\nu}- \delta\mN F
	  =
     2 I\frac{dF}{dI} \delta_{\mu 1}\delta^{\nu 1}
     		      		-F(I) \delta\mN;    \nn
	       I \eql -{\varphi'}^2  \e^{-2\alpha} < 0.         \label{e5}
\ear

    We will first obtain the general static, \cyl\ solution and then show
    that it cannot be \sol.

    The scalar field equation has the form
\beq                                                            \label{e6}
    \frac{1}{\sqrt{-g}}\d_{\mu}
       \biggl(\sqrt{-g}g^{\mn}\d_{\nu}\varphi\frac{dF}{dI} \biggr)=0
\eeq
    which, for $\varphi=\varphi(x)$,
    under the coordinate condition (\ref{harm}) is integrated to give
\beq
	\frac{dF}{dI} \varphi'(x) = C =\const.                 \label{e7}
\eeq
    Assuming that there is a known explicit expression for
    $F(I)$, one can find $I$ and $\varphi'$ as functions of $\alpha$.

    The ${1\choose 1}$ component of \eqs (\ref{e4}) can be written as
\bear
    {\alpha'}^2 - N^2 = 3\kappa\e^{2\alpha}
    				\biggl(2I \frac{dF}{dI}-F\biggr),\cm
		  N^2 \eqdef \thd(A^2 +AB + B^2) >0.     \label{e8a}
\ear
    Its integration gives
\beq                                                           \label{e9a}
    x = \pm \int d\alpha
              \biggl[ 3\kappa\e^{2\alpha}\biggl(F-2I\frac{dF}{dI}\biggr)
			+ N^2 \biggr]^{-1/2};
\eeq
    Reversing (if possible, explicitly) the dependence (\ref{e9a}), we
    obtain all unknowns as functions of $x$.

    Now, the following result is easily proved.

\Theorem {Proposition 5}
    {The system (\ref{e1}) does not admit a static, \cyl\ solution with a
    regular axis if the scalar field Lagrangian $F(I)$ has a linear weak
    field limit.}

    Indeed, let us use the regularity condition (\ref{Tfin});
    by (\ref{e5}) this means that both $|F(I)|$ and $|IF_I|$ are finite on
    the axis $x=\xa$ ($F_I\eqdef dF/dI$). Since $\gamma$ and $\xi$
    should be finite while $\beta\to -\infty$, we have
    $\e^{\alpha} \sim \e^{\beta} \to 0$. Meanwhile, it follows from
    (\ref{e7}) that $-IF_I^2 = C^2\e^{-2\alpha}\to \infty$. Thus
    simultaneously
\[
      |IF_I| < \infty  \quad {\rm and}\quad IF_I^2 \to \infty,
\]
    as $x\to\xa$, which is possible only if $I\to 0$ and $F_I\to \infty$.
    But this contradicts the assumed asymptotic linearity of the field
    theory which implies $F_I \to 1/2$ as $I\to 0$.

    We conclude that the Lagrangian (\ref{e1}) is unable to provide \sol\
    solutions.

    The above proof is quite similar to the one in \cite{c2} (see also
    \cite{NED}) where it was shown that solutions with a regular center
    cannot exist for nonlinear electrodynamics in \GR\ in the spherically
    symmetric case.  As in \cite{NED}, this proof is of local nature and
    does not depend on spatial asymptotics. Therefore, in particular,
    Proposition 5 is readily generalized to \GR\ with a cosmological
    constant.

\section   
	{Self-gravitating scalar field
         with the potential $V (\varphi)$ }

    Consider now a nonlinear field system with the Lagrangian
\beq
    L = \frac{R}{2\kappa} + \Half \varphi^{,\alpha}\varphi_{,\alpha}
	    - V(\varphi)                                      \label{e1'}
\eeq
    where $V(\varphi)$ is an arbitrary function. For the metric (\ref{e2})
    and $\varphi=\varphi(x)$, under the coordinate condition (\ref{harm})
    the Einstein equations take the form (\ref{e4}) with the EMT
\bear
    T\mN \eql \varphi_{,\mu}\varphi^{,\nu}                   \label{e5'}
     	- \delta\mN[\half \varphi^{,\alpha}\varphi_{,\alpha}
		    - V(\varphi)]      \nn
	  \eql
	  \half {\varphi'}^2 \e^{-2\alpha}\diag (1, -1, 1, 1)
			      + V(\varphi)\delta\mN.
\ear
    The scalar field equation is
\beq                                           		     \label{e6'}
    \frac{1}{\sqrt{-g}}\d_{\mu}
       \bigl[\sqrt{-g}g^{\mn}\d_{\nu}\varphi\bigr]+\frac{dV}{d\varphi}=0.
\eeq
    Since \eq (\ref{TTT}) holds as before, in the coordinates (\ref{harm})
    we again obtain \eqs (\ref{e7a}), reducing the behavior of the metric
    to one unknown $\alpha(x)$.
    The ${1\choose 1}$ component of \eqs (\ref{e4}) now gives
\bear                                                        \label{int}
    {\alpha'}^2 - N^2 \eql \fract{3}{2}\kappa{\varphi'}^2
					    -3\kappa V\e^{2\alpha},\nn
		  N^2 \al\eqdef\al \thd(A^2 +AB + B^2) >0
\ear
    (we take $N>0$ in agreement with \sect 2).
    On the other hand, a sum of \eqs (\ref{e4}) with
    the EMT (\ref{e5'}) and \eq (\ref{e6'}) give
\bear
    \alpha'' + 3\kappa V(\varphi) \e^{2\alpha} \eql 0,     	\label{e8}\\
    \varphi'' - (dV/d\varphi) \e^{2\alpha} \eql 0.              \label{e9}
\ear
    Thus the original set of equations has been reduced to (\ref{int}),
    (\ref{e8}) and (\ref{e9}), where \eq (\ref{int}) is a first integral of
    the other two.

    The following observation can be made directly from \eq (\ref{e8}):

\Theorem{Proposition 6}
    {In a \sol\ \cyl\ solution to the field equations due to
    (\ref{e1'}), the potential $V$ satisfies the condition
\beq
    \int_{-\infty}^{+\infty} V(\varphi(x)) \e^{2\alpha}dx =0.   \label{V0}
\eeq
    }

    Indeed, according to \sect 3.1, in a \sol\ solution one has
    $\alpha'\to N$ for both $x \to -\infty$ and $x\to +\infty$, therefore
    integration of (\ref{e8}) over $\R$ leads to (\ref{V0}).

    Proposition 6 means that \sol\ solutions can only be obtained with
    potentials having a variable sign.

    Let us now show a few ways of solving \eqs (\ref{int})--(\ref{e9}) by
    quadratures. The problem of solving the field equations with given
    $V(\varphi)$ is hard even in flat space --- see \eq (\ref{eqflat}).
    The general solution can be obtained in a few ways by specifying other
    functions involved.

    In practice, the quadratures and/or inverse functions, needed to
    express all quantities in a convenient way, are not available
    explicitly in most specific cases.  Therefore, being concerned with
    particular problems, it is useful to have various forms of the general
    solution at one's disposal.

\subsection*{General solution I: $x$-parametrization}

    The simplest parametrization of the general solution to \eqs
    (\ref{int})--(\ref{e9}) is obtained by specifying the function
    $\alpha(x)$. Indeed, from (\ref{e8}) one then finds $V(\varphi(x))$
    and after that $\varphi'(x)$ from (\ref{int}), which yields $\varphi(x)$
    by quadrature, so that the function $V(\varphi)$ is obtained in a
    parametric form. It is made explicit if one resolves $\varphi(x)$ with
    respect to $x$.

    Regularity of the solution on the axis is provided by $\alpha(x)$
    satisfying the condition (\ref{ax-x}), while to have a regular
    asymptotic one should fulfil \eq (\ref{as-x}).

\subsection*{General solution II: $\alpha$-parametrization}

    Introducing the notations
\beq
    U(\alpha) = 3\kappa V(\varphi)\e^{2\alpha},   \cm       \label{a1}
    y(\alpha) = {\alpha'}^2,
\eeq
    we can bring \eqs (\ref{int}) and (\ref{e8}) to the form
\bear
							    \label{a2}
    \frac{3\kappa}{2} \varphi_\alpha^2 \eql 1 - \frac{1}{y}(N^2 - U),\\
    y_\alpha \eql -2U(\alpha),                              \label{a3}
\ear
    where the subscript $\alpha$ means $d/d\alpha$. \eq (\ref{e9}) holds as
    their consequence.

    Now, if the function $U(\alpha)$ is specified, $y(\alpha)$ is found by
    quadrature from (\ref{a3}) and then $\varphi(\alpha)$ from (\ref{a2}).
    Furthermore, according to (\ref{a1}), $x(\alpha)$ is determined as
    follows:
\bear
	x \eql \pm \int d\alpha\Big/\sqrt{y(\alpha)}.    \label{a7}
\ear
    Thus all unknowns are expressed in terms of $\alpha$, and it remains to
    resolve the dependence $x(\alpha)$ with respect to $\alpha$ in order to
    express them in terms of $x$. To find $V(\varphi)$ it is also necessary
    to resolve $\varphi(\alpha)$ with respect to $\alpha$.

    Consider now the regularity conditions. On a regular axis $x\to
    -\infty$, in addition to (\ref{ax-x}), one obtains
\beq                                                       \label{ax1}
    \varphi = \varphi_{\rm ax} + O(\e^{Nx}); \cm U= O(\e^{2Nx}),
\eeq
    where $\varphi_{\rm ax}=\const$. These conditions imply the finiteness
    of both $\varphi$ and $V(\varphi)$ on the axis. The local flatness
    on the axis is provided, as before, by \eq (\ref{Nax}).

    To have a regular asymptotic, one should necessarily provide $V =
    o(\e^{-2\alpha})$ and hence $U(\alpha) = o(1)$ as $\alpha\to \infty$.

    The necessary condition (\ref{V0}) for a \sol\ nature of the solution
    has an analogue in terms of $\alpha$, also obtained from (\ref{e8}):
\beq
	\int_{-\infty}^{\infty} U(\alpha)d\alpha =0.         \label{U0}
\eeq
    If (\ref{U0}) holds, one can adjust the emerging integration constants
    to satisfy (\ref{Nax}) and (\ref{defect}) with given $\mu$.

\subsection*{General solution III: $\alpha$-parametrization}

    Substituting $U(\alpha)$ (defined in (\ref{a1})) from (\ref{e8}) into
    (\ref{e6'}) and treating $\varphi(\alpha)$ as a known function, one
    obtains a linear first-order differential equation for the unknown
    $y(\alpha) \equiv  {\alpha'}^2$:
\beq
    y_\alpha - y (2-3\kappa\varphi_\alpha^2) = -2N^2.           \label{b1}
\eeq
    Its solution and the respective expression for $U(\alpha)$ are
\bear
    y(\alpha) \eql -2N^2 \e^{2\alpha-\Psi}                      \label{b2}
    			\int \e^{\Psi -2\alpha}d\alpha,\\
    U(\alpha) \eql N^2                                          \label{b3}
    	  \biggl[ 1 + (2-3\kappa\varphi_\alpha^2)
                \e^{2\alpha-\Psi}\int \e^{\Psi - 2\alpha}d\alpha \biggr],
\ear
    where $\Psi (\alpha) = 3\kappa \int \varphi_\alpha^2\, d\alpha$.
    \eq (\ref{b2}) expresses $\alpha'$ in terms of
    $\alpha$, whence one finds by integration $x=x(\alpha)$ and consequently
    all unknowns as functions of $x$. As in solution I, to obtain the
    function $V(\varphi)$ it is necessary to resolve the dependence
    $x(\alpha)$ with respect to $\alpha$.

    As is easily verified, regularity on the axis takes place if and only if
\beq
    \varphi_\alpha = O(\e^{\alpha}) \ \then \
      \Psi = \const + O(\e^{2\alpha}) \cm {\rm as} \quad
		\alpha \to -\infty,                             \label{b4}
\eeq
    and in this case
\beq
    y = {\alpha'}^2 = N^2 + O(\e^{2\alpha}),                    \label{b5}
\eeq
    which describes $\alpha(x)$ near the axis more precisely than
    (\ref{ax1}).

    On the other hand, a necessary condition of a regular asymptotic is
    that, as $\alpha \to \infty$,
\beq
      \varphi' = o (1/x), \cm \mbox{so that}\cm
      \Psi \to \const.		                             \label{b6}
\eeq

\subsection*{General solution IV: $\varphi$-parametrization}

    Returning to \eqs (\ref{int})--(\ref{e9}), let us now put
    $\alpha=\alpha(\varphi)$. Then (\ref{int}) gives
\beq                 \label{f1}
    {\varphi'}^2 = \frac{N^2 - 3\kappa V(\varphi)\e^{2\alpha}}
     				{\af^2 - 3\kappa/2}.
\eeq
    where the subscript $\varphi$ denotes $d/d\varphi$.  Expressing
    $\varphi''$ from (\ref{f1}) and comparing it with (\ref{e9}), we arrive
    at a linear equation with respect to $V(\varphi)$:
\bearr
    V_\varphi + P(\varphi) V +  Q(\varphi)=0,
\nnn \cm                                           \label{f2}
    P(\varphi) = \frac{3\kappa}{\af} \biggl(1-
			\frac{\aff}{\af^2-3\kappa/2}\biggr), \cm
    Q(\varphi) = \frac{N^2 \aff
    				\e^{-2\alpha}}{\af(\af^2-3\kappa/2)}.
\ear
    Its solution for given  $\alpha(\varphi)$ is
\bear                                                          \label{Vphi}
    V(\varphi) \eql - \e^{-\int P d\varphi}
	\int \biggl( Q(\varphi)\e^{\int P d\varphi}\biggr)d\varphi,
\\
    \e^{-\int P d\varphi} \eql		                     \label{f3}
		\biggl(\frac{3\kappa}{2\af^2}-1\biggr)
				\e^{-3\kappa\int d\varphi/\af}.
\ear
    With known $\alpha(\varphi)$ and $V(\varphi)$, from (\ref{f1}) one
    finds $x(\varphi)$ and, reversing it, determines all unknowns as
    functions of $x$, thus completing the solution.

    Regularity conditions for $\alpha(\varphi)$ are found from those for
    $\varphi(\alpha)$ in (\ref{b4}), (\ref{b6}). Thus, a regular axis may
    be provided by
\beq
      \e^\alpha \sim |\varphi-\varphi_{\rm a}|^k, \cm  k\leq 1, \label{f4}
\eeq
    where $\varphi_{\rm a}$ is the (finite) value of $\varphi$ on the axis,
    while a similar condition for a regular asymptotic is
\beq
     \e^{-\alpha} \sim |\varphi-\varphi_\infty|^k, \cm  k < 1.   \label{f5}
\eeq

\section{Flat-space limit}  

    One can naturally expect that a solution to the nonlinear scalar
    field equation in Minkowski space will be obtained
    from a solution for a self-gravitating field in the limit $\kappa\to 0$.
    This is indeed the case, but the transition should be performed with
    certain care. Indeed, as $\kappa\to 0$, matter decouples from gravity,
    therefore, generically, the metric from a solution with self-gravitating
    matter should tend to a vacuum solution of \GR\ with the corresponding
    symmetry and only under additional assumptions it will tend to flat
    metric. Moreover, when it happens, the matter in the same limit may
    acquire different forms depending on how the parameters of the original
    solution (e.g., the integration constants) depend on $\kappa$, and this
    makes a separate assumption, therefore the same self-gravitating
    solution can pass to different flat-space ones. It thus seems
    instructive to follow this limit for our solutions I--IV.

    Let us require that, as $\kappa\to 0$, the metric tend to the
    Minkowski metric in cylindrical coordinates satisfying (\ref{harm}),
    namely,
\beq
	ds^2 = dt^2 - \e^{2x}dx^2 - dz^2 - \e^{2x} d\phi^2.  \label{E4}
\eeq
    The conventional form of the Minkowski metric in cylindrical
    coordinates is recovered by putting $\e^x = r$.

    The scalar field and the potential $V$ should, in the same limit,
    satisfy the flat-space equation
\beq
    \varphi'' - (dV/d\varphi) \e^{2x} =0.              	     \label{eqflat}
\eeq

    Let us put for simplicity $A=B=N=1$ in the self-gravitating
    solutions themselves, although one might, in general, only require
    $A(\kappa)\to 1$, $B(\kappa)\to 1$, $N(\kappa)\to 1$ as $\kappa\to 0$.

\para{Solution I.}
    Flat space is obtained by specifying $N=1$ and $\alpha\equiv x$, to
    which one can proceed from a given function $\alpha(x)$ along any
    sequence of functions, parametrized by $\kappa\to 0$. Then
    \eqs (\ref{int}) and (\ref{e8}) are evidently satisfied for $\kappa=0$,
    while (\ref{e9}) takes the form (\ref{eqflat}).

\para{Solution II,} (\ref{a2})--(\ref{a7}).
    Denoting, in accordance with (\ref{a1}),
\beq
    \int U(\alpha)\, d\alpha = 3\kappa X(\alpha) -C       \label{X}
\eeq
    where $X(\alpha)$ is a $\kappa$-independent function and $C$ is
    an integration constant, one can rewrite \eq(\ref{a3}) in the form
\beq
    \varphi_{\alpha}^2 = \frac{2}{3\kappa}                    \label{X1}
	   	\frac{2C-1 -6\kappa X + 3\kappa V\e^{2\alpha}}
	   	     {2(C-3\kappa X)}
\eeq
    whence it follows that to have a proper limit we must put $C= 1/2$ and
    consequently, in the limit $\kappa\to 0$,
\beq
    \varphi_{\alpha}^2 = 2(V\e^{2\alpha} - 2X(\alpha)).    \label{int-fl}
\eeq

    In the metric (\ref{E4}) $\alpha\equiv x$, and it is directly verified
    that, under the substitution $\alpha =x$, \eq (\ref{int-fl}) is a first
    integral of (\ref{eqflat}).

    It remains to provide a proper transition in the metric. To this end, it
    is sufficient to put in (\ref{e7a}) $A=B=1$, so that $N^2=1$, and to
    verify the coincidence of $x$ and $\alpha$ in (\ref{a7}) for
    $\kappa=0$. With (\ref{X}) and $C=1/2$, \eq (\ref{a7}) is
    (omitting $\pm$) rewritten as
\beq
    x = \int d\alpha\,[1- 6\kappa X(\alpha)]^{-1/2},        \label{x-fl}
\eeq
    so that for $\kappa=0$ one obtains $x = \alpha$ up to an inessential
    additive constant.

\para{Solution III,} (\ref{b2})--(\ref{b3}).
    By definition, $\Psi\sim \kappa$ as $\kappa\to 0$, and from
    (\ref{b2}) one finds up to $O(\kappa)$:
\beq
    y\equiv {\alpha'}^2 =                                     \label{ylim}
    	1 - \Psi - 2\e^{2\alpha}\int\e^{-2\alpha}\Psi\,d\alpha,
\eeq
    whence it follows that, first, $\alpha= x + O(\kappa)$ (up to an
    additive constant) and, second, according to (\ref{b4}),
\bear
    U \equiv 3\kappa \e^{2\alpha} V                           \label{Ulim}
	\eql \half \Psi_\alpha + \Psi
			+ 2\e^{2\alpha}\int\e^{-2\alpha}\Psi\,d\alpha\\
							      \label{Vlim}
	\eql 3\kappa\e^{2\alpha}
             \int\e^{-2\alpha}\varphi_\alpha\varphi_{\alpha\alpha} d\alpha.
\ear
    The expression (\ref{Vlim}) is obtained from (\ref{Ulim}) by twice
    integrating by parts.

    Now, the flat-space equation (\ref{eqflat}) leads to the following
    expression for $V(\varphi)$ for given $\varphi(x)$:
\beq
      V = \int\e^{-2x}\varphi'\varphi'' dx.                   \label{Vflat}
\eeq
    Evidently (\ref{Vlim}) yields (\ref{Vflat}) since in the same
    limit $\alpha$ coincides with $x$.

    One should note that, given the same functional dependence
    $\varphi(\alpha)$ in (\ref{b3}) and $\varphi(x)$ in (\ref{Vflat}),
    the resulting functions $V(\varphi)$ will be, in general, different.

\para{Solution IV,} (\ref{Vphi})--(\ref{f3}).
    Suppose that $\alpha(\varphi)$ is $\kappa$-independent and $V(\varphi)$
    does not blow up as $\kappa\to 0$.
    Then, in this limit, \eq (\ref{f1}) turns into the equality
    $\alpha_\varphi^2 {\varphi'}^2 = 1$, whence it follows $\alpha=x$
    (leading to the metric (\ref{E4})) for a proper choice of the sign and
    origin of $x$.

    Furthermore, in (\ref{f2}) $P(\varphi) \to 0$ as $\kappa\to 0$, while
    the limiting form of $Q(\varphi)$, with $N=1$ and $\alpha=x$, is
\beq                                                               \label{Q}
    Q(\varphi) = x_{\varphi\varphi}\e^{-2x} /x_\varphi^3\cm (\kappa\to 0).
\eeq
    Therefore the solution of (\ref{f2}), namely,
    $V(\varphi)= -\int Q(\varphi) d\varphi$, coincides with that of
    (\ref{eqflat}) with respect to $V(\varphi)$ for known $x(\varphi)$.
    As in solution II, the limiting function $V(\varphi)$ is different from
    the one in \eq(\ref{f3}).

    Thus correct transitions from self-gravitating to flat-space solutions
    have been obtained for all four forms of the general solution.

\section {Examples}

\para{1.}
    The first example concerns the choice of $V(\varphi)$ when \eqs
    (\ref{int})--(\ref{e9}) are integrated directly --- the Liouville
    potential. Let us denote
\beq                                                           \label{Wpsi}
	\kappa V = 2W(\psi), \cm \sqrt{\kappa/2}\varphi = \psi
\eeq
    and choose
\beq                                                           \label{Liu}
       W (\psi) = W_0 \e^{\lambda\psi}, \cm W_0, \lambda = \const.
\eeq
    Then \eqs (\ref{e8}), (\ref{e9}) combine to give
\beq                                                           \label{la}
	(6\psi + \lambda\alpha)'' =0, \cm 6\psi' + \lambda\alpha' =C,
\eeq
    where the integration constant $C$ should be equal to $\lambda N$
    with $N>0$ introduced in (\ref{ABN}) if we require a regular axis at
    $x\to -\infty$ (where $\psi'\to 0$ while $\alpha' \to N$). Then
    (\ref{int}) takes the form
\beq                                                           \label{int1}
    \alpha'{}^2 - \frac{\lambda^2}{12}(N-\alpha')^2
	= N^2 - 6W_0 \e^{\lambda^2Nx/6 + 2\alpha(1-\lambda^2/12)}.
\eeq
    In case $\lambda^2 = 12$ its integration gives
\beq
    \alpha = Nx - (3W_0/2N^2) \e^{2Nx} + \const.               \label{a12}
\eeq
    For $\lambda^2 \ne 12$ one obtains
\bearr
    \e^{k\alpha} = \e^{(k+1)Nx} \frac{\sqrt{6|kW_0|}}{2N}
		\biggl[ \e^{N(x+x_0)} - \eps \e^{-N(x+x_0)}\biggr],
\nnn                                                           \label{a_12}
     k \eqdef \lambda^2/12 -1, \cm \eps \eqdef \sign(kW_0).
\ear
    This completes the integration. It is easily confirmed that the
    asymptotic regularity condition $\alpha\sim Nx$ as $x\to \infty$ cannot
    be fulfilled, in agreement with the above Proposition 6.

\para{2.}
    In the general case the potential $V(\varphi)$ may be
    $\kappa$-dependent, and this opportunity can be used for obtaining
    specific solutions for self-gravitating scalar fields corresponding to
    known flat-space solutions. In particular, if, for given $V(\varphi)$
    in flat space, $\varphi(x)$ and hence $V(x) = V(\varphi(x))$ are known,
    the same dependence ascribed to $V(\alpha)$ should lead to a
    self-gravitating solution in $\alpha$-parametrization, with a certain
    function $V(\varphi,\kappa)$ which tends to the original potential
    $V(\varphi)$ as $\kappa\to 0$. A reason is that, under the coordinate
    condition (\ref{harm}), $\alpha = x$ in the flat-space metric. But, as
    was already clear from our general consideration, such a transition
    only occurs under special assumptions on the $\kappa$-dependence of the
    integration constants.

    Consider, e.g., in flat space-time
\beq                                                          \label{V2-fl}
    V(\varphi) = \lambda \varphi^{2n},
    	\cm  \lambda = \const > 0,\cm n =\const\ne 1.
\eeq
    Then the flat-space scalar field equation (\ref{eqflat}) reads
\beq
      \varphi'' = 2n\lambda \varphi^{2n-1} \e^{2x}            \label{ef2-fl}
\eeq
    and has a special solution of the form
\beq
    \varphi = \varphi_0 \e^{\nu x},                           \label{f2-fl}
      \cm   \nu = \frac{1}{1-n}, \cm
      \varphi_0 = \biggl(\frac{2n\lambda}{\nu^2}\biggr)^{\nu/2}.
\eeq
    The potential $V(\varphi)$ is expressed in terms of $x$ as
\beq                                                          \label{V2-x}
     V(\varphi) = \lambda \varphi_0^{2n}\e^{2nx/(1-n)}.
\eeq
    Now, we can seek a self-gravitating solution with a proper flat-space
    limit assuming
\beq                                                          \label{V2}
     V(\varphi) = \lambda \varphi_0^{2n}\e^{2n\alpha/(1-n)},
\eeq
    with some constants $n$ and $\varphi_0$.
    Integrating according to (\ref{a1})-(\ref{a7}), we obtain:
\bear
    \e^{\alpha/(1-n)} = \frac{\sqrt{C}}{\cosh [C_1(x-x_0)]},   \label{al-2}
\ear
    where the constant $C$ arises from integration in (\ref{a3}) and
    $C_1 = 3\kappa\lambda\sqrt{C}\varphi_0^{2n}$. Further integration gives
\beq
    \e^{\alpha/(1-n)}                                          \label{af-2}
	= \sqrt{C} \sin \biggl[\frac{1}{n}
		\sqrt{\frac{3\kappa}{2}}(\varphi-\varphi_1)\biggr]
\eeq
    with one more integration constant $\varphi_1$, so that the potential
    takes the form
\beq
    V(\varphi) = \lambda\varphi_0^{2n}\biggl\{                \label{Vf-2}
		\sqrt{C}\sin\biggl[
      \frac{1}{n}\sqrt{\frac{3\kappa}{2}}\ (\varphi-\varphi_1)
			\biggr]\biggr\}^{2n}
\eeq
    which resembles the well-known sine-Gordon nonlinearity.
    This function tends to (\ref{V2-fl}) as $\kappa\to 0$ if and only if
    the constants $\varphi_1$ and $C$ depend on $\kappa$ in such a way
    that $\varphi_1\to 0$ and $\varphi_0\sqrt{3\kappa C/2} \to n$.

\para{3.}
    Let us try to construct a \sol\ configuration, again on the basis of
    general solution II, \eqs (\ref{a2})--(\ref{a7}).  Putting, in
    accordance with the regularity conditions,
\beq
     y(\alpha) = N^2\biggl(1 + \frac{A}{\cosh^2 n\alpha}\biggr),\label{y-3}
\eeq
    where $A>-1$ and $n>0$ are constants, one obtains from (\ref{a3})
\beq
      U(\alpha) = N^2 An \frac{\sinh n\alpha}{\cosh^3 n\alpha}. \label{U-3}
\eeq
    It is an odd function, manifestly satisfying (\ref{U0}); it vanishes
    sufficiently rapidly as $\alpha\to\pm \infty$ if $n\geq 1$.

    The integral in (\ref{a7}) is easily found, leading to the following
    relation between $x$ and $\alpha$:
\beq
   \sinh n\alpha = \frac{1}{\sqrt{A+1}}\sinh (Nnx+\ln\sqrt{A+1}),\label{xa-3}
\eeq
    with a correct behavior at large $\alpha$ and $x$. It remains to find
    $\varphi$. Let us take for simplicity $N=1$ (a flat rather than string
    asymptotic).  \eq (\ref{a3}) gives
\beq
    \frac{3\kappa}{2}\ \varphi_\alpha^2 =                       \label{fa-3}
    \frac{A(\cosh n\alpha + n \sinh n\alpha)}
    	 {\cosh n\alpha (A + \cosh^2 n\alpha)}.
\eeq
    The r.h.s. is positive for all $\alpha\in \R$ under the conditions
    $A>0$, $n\leq 1$. Recalling that $n\geq 1$ by the regularity
    requirement, we are left with $n=1$. Thus we should put $n=1$ in
    (\ref{y-3})--(\ref{xa-3}), and the following expression for $\varphi$
    is obtained:
\beq
    \sqrt{\frac{3\kappa}{2}} \ \varphi =                         \label{f-3}
	\sqrt{A} \int \e^{\alpha/2}
		[\cosh\alpha (A + \cosh^2\alpha)]^{-1/2}\,d\alpha,
\eeq
    which can be written in terms of elliptic functions. Indeed, putting
    $\e^{2\alpha} = u$, one obtains
\beq
    \sqrt{\frac{3\kappa}{2}} \ \varphi =                         \label{f-4}
       \sqrt{2A}\int \frac{du}{(u+1)^{1/2}[u^2+ 2A(A+1)u +1]^{1/2}},
\eeq
    whence (integrating from finite $u$ to infinity)
\beq
    \varphi = \frac{2}{\sqrt{3\kappa}}
	      \biggl(\frac{A}{A+1}\biggr)^{1/4}
	      F\Biggl(
	       \arcsin \sqrt{\frac{4B}{\e^{2\alpha}+2A+2B}},\
	       \sqrt{\frac{A+B}{2B}}\Biggr),
\eeq
    where $F$ is the elliptic integral of the first kind and
    $B=\sqrt{A(A+1)}$. It is easy to verify that all the regularity
    conditions are satisfied; in particular, $\varphi \sim \e^{-\alpha}$ as
    $\alpha\to\infty$ and $\varphi$ tends to a finite limit as
    $\alpha\to -\infty$.

\para {Acknowledgment.}
    We acknowledge partial financial support of Russian Ministry of
    Industry, Science and Technologies and Russian Ministry of Education.

\end{document}